# Topological Extraordinary Optical Transmission

[1]K. Baskourelos, [2]O. Tsilipakos, [3]T. Stefański, [4]S. F. Galata, [2,5]E. N. Economou, [2,6]M. Kafesaki, and [1]K. L. Tsakmakidis

*[1]Section of Condensed Matter Physics, Department of Physics, National and Kapodistrian University of Athens, Panepistimioupolis, GR-157 84 Athens, Greece*
*[2]Institute of Electronic Structure and Laser, Foundation for Research and Technology Hellas, Heraklion, Crete GR-70013, Greece*
*[3]Gdańsk University of Technology, Faculty of Electronics, Telecommunications and Informatics, ul. G. Narutowicza 11/12, 80-233 Gdańsk, Poland*
*[4]Department of Electrical and Electronic Engineering, University of West Attica, Athens, Greece*
*[5]Department of Physics, University of Crete, GR-70013 Heraklion, Crete, Greece*
*[6]Department of Materials Science and Technology, University of Crete, GR-70013 Heraklion, Crete, Greece*

To whom correspondence should be addressed: ktsakmakidis@phys.uoa.gr

**Abstract:** The incumbent technology for bringing light to the nanoscale, the near-field scanning optical microscope, has notoriously small throughput efficiencies – of the order of $10^{-4}$ - $10^{-5}$, or less. We report on a *broadband*, topological, unidirectionally-guiding structure, not requiring adiabatic tapering and *in principle* enabling near-*perfect* (~100%) optical *transmission* through an *unstructured single* (POTUS) arbitrarily-subdiffraction slit at its end. Specifically, for a slit width of just $\lambda_{eff}$/**72** ($\lambda_0$/138) the attained normalized transmission coefficient reaches a value of **1.52**, while for a unidirectional-only (non-topological) device the normalized transmission through a $\lambda_{eff}$/21 (~$\lambda_0$/107) slit reaches **1.14** – both, limited only by inherent material losses, and with zero reflection from the slit. The associated, under ideal (ultralow-loss) conditions, near-*perfect* optical *extraordinary transmission* (POET) has implications, among diverse areas in wave physics and engineering, for high-efficiency, maximum-throughput nanoscopes and heat-assisted magnetic recording devices.

A so-far unsolved challenge for many important photonic applications relying on nanofocusing, such as all-optical data writing/storage, heat-assisted magnetic recording (HAMR), nanoimaging, spectroscopy, sensing, near-field scanning optical or thermal nanoscopy, thermal scanning probe lithography, and nanoscale thermometry, is to focus, *with high efficiency* ~100 µW of energy to a ~10 nm (or less) spot on a planar surface [1-5]. This is an extremely large light intensity, hundreds of millions of times larger than, e.g., the intensity of sunlight on the surface of the earth. This power density is ~1.000x the throughput of gold-coated tapered optical fibers used in Near-field Scanning Optical Microscopes (NSOMs), which is the incumbent technology allowing the focus of light on the nanoscale [1, 2]. Typical optical transmission efficiencies of NSOM probe tips are between ~$10^{-4}$ – $10^{-5}$ (or less), while the minimum optical throughput efficiency required for commercial applications to be powered by inexpensive 10 mW diode lasers is ~1%. Conventionally, focusing of light is performed in the far field with lenses. The shortest-wavelength light-source (among those that could potentially be produced in billions of units), a 400 nm laser diode, offers a ~200 nm minimum diameter spot in the far field in air (with an infinite aperture). To reach a ~10 nm (or less) spot in the far-field, we would require an index of refraction larger than 7, which is not available with typical natural materials. A similar constraint is usually found with guided modes in dielectric waveguides, with them too yielding a spot size far larger than the required ~10 nm (or less) spot size [1-5].

To overcome the above formidable challenges, and to see how (for an ideal loss-less structure) near-*p*erfect *o*ptical *e*xtraordinary *t*ransmission (POET) through deep-subdiffraction slits can arise, we here consider **two** classes of structures; **first**, one which is only unidirectional but non-topological and, **second**, a unidirectional but also truly topological [11] structure. In the first case (see Fig. 1a), a surface magnetoplasmon (SMP) exists at the interface between a magnetically biased gyroelectric semiconductor (InSb) and silicon (Si) [6-9, 16], whereas in the second case the Si layer is replaced by a semi-infinite plasmonic, negative permittivity material (e.g. a polaritonic material at THz frequencies [17], or a metal such as Ag in the near-IR and optical region), directly touching the InSb layer (also semi-infinite in this case) (see Fig. 1b) [10, 11]. Note that in the first case the substrate and superstrate need to be "truncated" by a good conductor (PEC in the simulations) to avoid back-reflection via bulk modes [7]. In both cases, a complete unidirectional propagation (**CUP**) frequency band opens up, inside which the SMP can only propagate in one direction (towards the right, in Fig. 1), owing to breaking of time-reversal symmetry by the applied magnetic field bias (*B*), leading to the absence of propagation in the 'negative' (towards the left) direction.

The reason for deploying the second (topological [11]) structure is because nonreciprocity is not on its own always sufficient to give rise to genuinely unidirectional surface modes. Specifically, if nonlocal effects (spatial dispersion) in the plasmonic or semiconductor material(s) cannot be neglected (which is usually true for tight light localization and focusing, as in the present case) the CUP region in the band-diagram may close, destroying the herein desirable one-way property [8-11]. To overcome the detrimental role of nonlocality, truly topological (rather than 'only' unidirectional) structures are required, such as the second structure (Fig. 1b) [10, 11]. Here, *immunity to nonlocal effects is warranted by the fact that within the CUP band the longitudinal wavevector $\beta$ need not diverge (cf. Figs. 1b & 1a) – hence, as Ref. [11] has shown, nonlocal effects (usually prominent for large wavevectors) are not harmful for such finite-$\beta$ structures.* Besides, unidirectionality occurs within the opening of a bulk-mode bandgap, and is associated with a *topological invariant number* related to the bulk excitations – usually an integer number conserved under continuous deformations that do not close the bandgap, i.e., with the Chern number, which can simply be seen as the number of windings for a modal evolution in the momentum space [9-11]. Indeed, after a full evolution in momentum space, a mode does not in general always return exactly back to its original form but may instead acquire an additional phase term, the Berry phase, equal to $2\pi C$, where C is the aforementioned Chern number for our given mode. Crucially, this winding owing to the modal evolution is *completely immune to perturbations* that are not large enough to destroy the bandgap. Consequently, when two media sharing the same bandgap, but having different topological properties, are brought together, the aforementioned bandgap will necessarily close at the two-media interface in order to facilitate the change in the topological invariants of the two media, thereby giving rise to *truly topological*, unidirectional surface waves (SMPs here) at that interface. These waves are immune even to nonlocal plasmonic effects, and their dispersion extends over the full bandgap [9-11]. Furthermore, the 'bulk-edge correspondence' principle ensures that the total number of unidirectional surface waves at the two-media interface is equal to the difference between the 'gap Chern' numbers of the two media (the gap Chern number is the summation of the Chern numbers of all the modes below the bandgap). For our topological structure, formed by the interface between a magnetized plasma and an plasmonic ($\varepsilon_r < 0$) material, the difference between the two gap Chern numbers is exactly equal to unity – and, as a result, *exactly one unidirectional* surface wave extends across (closes) the bandgap [11], which is ideal for what we are looking for in the present context.

Full-wave (COMSOL) simulations reveal that a source with a spectrum within the CUP band (cf. Fig. 2a) launches all the incoming SMP wave energy, without reflection, to a localized surface state existing at either the Si-InSb interface (first structure) or at the InSb-[plasmonic material] interface (second structure) (see Figs. 2c, d) – with that localized surface state then decaying with time owing to dissipative propagation losses arising from the InSb and plasmonic layers. Importantly, because the structure is inherently unidirectional within the CUP band, there is no need for adiabatic tapering, unlike NSOM structures, where such a tapering is needed to prevent back-reflections – but which, as a result, *unavoidably leads to elongated lengths and to increased propagation losses*. Thus, our two structures can have an extremely short propagation length, leading to increased throughput efficiencies and more compact designs.

Having outlined the nonreciprocal or topological character of the two structures, next, we move on to reveal a unique practical application they can ideally facilitate, namely near-perfect wave focusing enabled via *near-perfect transmission through deep-subdiffractional* slits (under ultralow-loss conditions). Nonlocal effects are not considered in the analysis below, thus it is to be understood that robust performance as described by the theory is only attained by the second (topological) structure.

Consider the two structures described in Fig. 1 with a deep-subdiffractional slit opened at their ends. If the incident SMP pulse carries power $|s_0|^2$, then, in an approximative analysis [8], the amplitude $a_q$ of an excited $q$ mode within the $\Delta\omega$ band, localized at the terminating interface, will vary with time as [12, 13] $da_q/dt = j\omega_q a_q - (\gamma_R + \gamma_T + \gamma_0)a_q + \kappa s_0$, where in general $\gamma_R$ is the decay rate of the mode in the backwards direction (if there is any – see next) where the SMP is coming from, $\gamma_T$ the tunneling rate of the mode through the slit (in the forward direction), $\gamma_0$ the decay rate of the mode because of dissipative losses, and $\kappa$ is the in-coupling coefficient. Then, for the transmitted through the slit (in free space) field we have: $s_T = t_D s_0 + \sqrt{2\gamma_T} a_q e^{i\varphi}$, where $t_D$ is the direct transmission coefficient of the SMP through the slit, and $\varphi$ the phase difference between the in-coupling and out-coupling processes of the resonant mode. As a result (see details in the Suppl. Inform.), in general, the power transmission through the slit contributed by this mode at $\omega = \omega_q$, will be: $T = |s_T|^2/|s_0|^2 = |t_D + (\sqrt{2\gamma_T}\kappa e^{i\varphi}) / (\gamma_R + \gamma_T + \gamma_0)|^2$. Furthermore, because of energy conservation and broken time-reversal symmetry (absence of back-reflections) in these structures, we have (see Suppl. Inform.) $\gamma_R = 0$ and $\kappa^2 = 2\gamma_T$, from where we finally obtain the contribution of the $a_q$ mode (at $\omega = \omega_q \in \Delta\omega$) to the transmission through the deep-subdiffractional slit:

$$T_{a_q} \cong \frac{2\gamma_T}{2\gamma_T + \gamma_0}. \qquad (1)$$

In deriving the above relation, which illustrates the main physical content of the present study, i.e., the competition between absorption and tunneling rates in the POET process, it was also assumed that the direct transmission coefficient $t_D$ through the deep-subdiffraction slit is small (for deep subwavelength slits), $t_D \sim 0$, and that $\gamma_0 << 2\gamma_T$. An exactly similar relation to Eq. (1) holds for all $q$ modes throughout the continuous unidirectional frequency region in the CUP bands of Figs. 1, 2. Thus, as long as $\gamma_0 << 2\gamma_T$, or if $\gamma_0 \rightarrow 0$ (very small but finite losses, e.g., similarly to 'perfect' focusing in negative-index lenses [14, 15]), the transmission through the slit can approach unity – irrespective of how subdiffractional the slit is.

Equation (1) has a precise **physical meaning**: Because in these unidirectional or topologically-protected structures the back-reflection channel is completely eliminated, the incident SMP can now remain close to the slit for much longer times (if losses are small), therefore it also has more time to slowly tunnel through it – a feat that fundamentally does not occur in standard (bi-directional) configurations because therein an incident guided wave is always back-reflected from a termination. Thus, as long as we design a structure to be strictly unidirectional, and allow for larger incident-wave lifetimes ($\sim\gamma_0^{-1}$) close to the slit compared to the tunneling time ($\sim\gamma_T^{-1}$), i.e., have a low-loss structure, almost *all* the wave energy will eventually escape through the slit, as it has nowhere else to 'go' during $\gamma_0^{-1}$ but to *tunnel* through the slit (the localized wave is a *surface* magnetoplasmon). We may thus target spectral regimes for which dissipative losses are small, or we can reduce the temperature of the device's operation. Unlike what is conventionally ascertained, even within the field of extraordinary optical transmission EOT [1, 2], here, in principle, there is no fundamental maximum-efficiency limit that can be attained for the transmission through a subdiffractional slit. In a somewhat similar manner to the distinction between 'internal' (intrinsic) and 'external' (extrinsic) quantum efficiency in light-emitting devices (LEDs) or solar panels, here the 'intrinsic' nanofocusing efficiency of the structure is close to unity, and for practical applications the extrinsic nanofocusing efficiency (tunneling time) [17, 18] and the precise outcoupling design can be optimized with different designs and operational frequency regimes.

The potential benefits of the present scheme compared to existing EOT and subdiffraction wave-focusing devices become particularly noticeable in the deep-subdiffraction regime (e.g., slit dimensions in the $\lambda_0/100$ regime, or smaller) where the throughput efficiencies of standard devices are vanishingly small [1, 2]. Furthermore, compared to negative-refractive-index wave-focusing media, which they too in principle allow for 100% focusing efficiencies even deep below the diffraction limit but are narrowband [14], our present broadband scheme – relying on natural, magnetized-semiconductor materials – does not have issues with defining three-dimensional *homogenized* effective refractive indices, or with spatial dispersion (for the topological structure), or with surface roughness and structural imperfections / inhomogeneities, i.e. it is *stable* in the optogeometric-parameters space. Furthermore, the structure need not be completely lossless; for any given $\gamma_0$, the extrinsic outcoupling efficiency may be designed such that $\gamma_T \gg \gamma_0/2$ [18, 19].

Figures 3 and 4 report how, for both structures, the throughput varies with different slit vertical positions (offsets) (Fig. 3) or different frequencies within the CUP band (Fig. 4). For the nonreciprocal structure, a transmission as high as $T \sim$ **23%** is found through extremely small slit sizes, of the order of $\lambda_0/107$ ($\sim \lambda_{\text{eff}}/21$, where $\lambda_{\text{eff}}$ is the effective wavelength for the excited guided mode, also denoted as $\lambda_g$) (Fig. 3a). The transmission reaches (not shown) $T \sim 18\%$ through a slit of size $\lambda_0/214$ ($\sim \lambda_{\text{eff}}/42$). The afore-reported transmission is equivalent, as is usually done in EOT studies [1, 2], to normalized (to the incident in-the-slit-only power) transmission *greater than unity*, namely $T_{\text{norm}} \sim$ **1.14**, i.e. it is a truly 'extraordinary' transmission [1]. The dependence of the transmission for this device on both the slit offset (see upper left inset in Fig. 3(a)) and size, are presented in Fig. 1S in the Supplementary Information. One the other hand, for the even more robust topological structure, the transmission through an even smaller slit, of size $\boldsymbol{\lambda_{\text{eff}}/72}$ (= $\lambda_0/138$, $n_{\text{eff}} = 1.913$, Fig. 3b), reaches **33%** at the frequency of 2.18 THz (cf. Fig. 4d). This means that the *normalized* (to the incident in-the-slit-only power) transmission through this slit is **1.52**, that is, it is 'extraordinary', too [1].

These throughput efficiencies are, for given losses, the maximum possible since the reflection is zero throughout the CUP region (cf. Figs. 4b, e); they are here still not ~100% owing to inherent material losses, sub-optimal external (extraction) efficiencies, as well as because of purely numerical reasons: The localized fields in these structures are ultra-singular [7-10], i.e., they require extremely fine numerical mesh-discretization close to the localization region. Exactly analogous situations are well-known to occur, e.g., for 'perfectly' focusing negative-refractive-index structures, where, there too, 'perfect' (that is, infinite) wave focusing cannot be observed numerically but only analytically [14, 15].

The maximum sustainable optical power at the output of our wave-focusing structures will, ultimately, be limited by self-heating near the slit, in light of the attained large field enhancements [7], necessitating either lower in-coupled wave powers, or materials and heat sinks with high thermal conductivity, or suitable cooling techniques [19]. Further realistic limiting factors, not considered here, are nonlocal effects, which nonetheless are not harmful for the truly topological structure and even for the nonreciprocal structure when moderate material losses (as in here) are considered. Further realistic effects should be plasmon enhanced blackbody radiation, which is expected to be high in light of the *enhanced density of states* at the localized topological resonance, and nanoscale phonon tunneling, which may overall cause structural deformations of the slit and thermal instabilities of nearby devices [1-5]. Furthermore, although our scheme here addresses the problem of how waves can efficiently be focused at deep-subdiffractional dimensions, its *temporal* resolution too can be addressed using standard techniques in NSOM and HAMR technologies – a requirement for high areal throughput in optical lithography applications according to the well-accepted, and industry adopted, Tennant's Law [20].

Our results establish that topology [6-11], which until now has primarily being studied in relation to its ability to provide protection (immunity) for *propagating* waves, has a hitherto unexplored but important role to play for *localized* waves, too – allowing for an efficient solution to a key problem plaguing wave physics and engineering, namely the keen difficulty to focus waves, with high efficiency, below the diffraction limit. Potential applications that could benefit from this new type of *broadband topological microscopy* include nanoscale sensors, maximum-throughput heat-assisted magnetic recording devices, near-field scanning optical or thermal nanoscopy, thermal scanning probe lithography, and nanoscale thermometry [1-5, 20].

## Acknowledgements

K. B. and K.L.T. were supported by the General Secretariat for Research and Technology (GSRT) and the Hellenic Foundation for Research and Innovation (HFRI) under Grant No. 1819.

# Figures

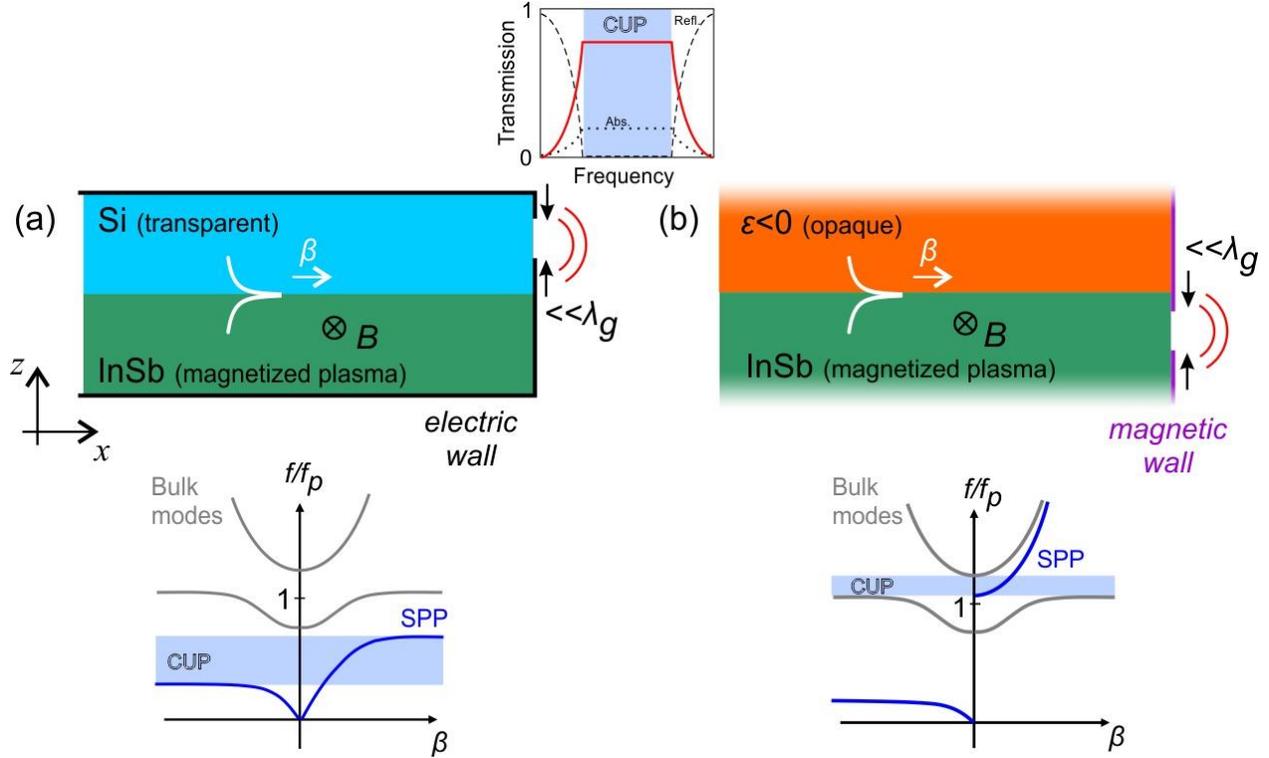

**Fig. 1.** Topological nanoscopy. Two different structures are considered: The first one (left side) is schematically shown in (**a**), and is 'only' nonreciprocal but not topological, thereby harmed by nonlocal effects (not considered here). The second structure (in **b**) is similar to the one shown on the left panel but without the Si layer, where now a semi-infinite InSb layer is directly interfaced with a plasmonic semi-infinite space. This second structure is truly topological [9-11], exhibiting robust unidirectionality even in the presence of nonlocal plasmonic effects. A light pulse, whose bandwidth is in the 'complete unidirectional propagation' (CUP) region excites a single *surface* magnetoplasmon (SMP) mode, reaching in both cases the waveguide end where there is a deep-subdiffraction slit (e.g., $\lambda_{eff}/72$, $\lambda_{eff}$ being the effective wavelength inside the guide). In these structures, the pulse cannot be back-reflected, hence it has sufficient time to tunnel through the slit, so long as the rate of dissipative losses is considerably smaller than the tunneling rate through the slit – leading to much increased throughput (transmission) efficiency through the subdiffractional slit; ideally (for a lossless structure, where absorption is negligibly small) approaching unity throughout the broadband CUP band (top panel). Symbols B, $f_p$, $\beta$, $\lambda_g$, denote, respectively, the biasing magnetic field, the plasma frequency of InSb, the propagation constant of the surface wave, and the effective wavelength of the surface wave, $\lambda_g = \lambda_{eff} = \beta/2\pi$. Note that the structure of panel (**a**) is closed at the top and bottom sides with an electric wall (black line) to avoid backwards bulk guided modes in Si. This is not needed for the structure of panel (**b**) where there is no propagation possibility in the bulk plasmonic material. In the case of structure **b** the slit is an opening in a magnetic wall, in order to avoid unwanted surface waves along the *z* axis at the termination.

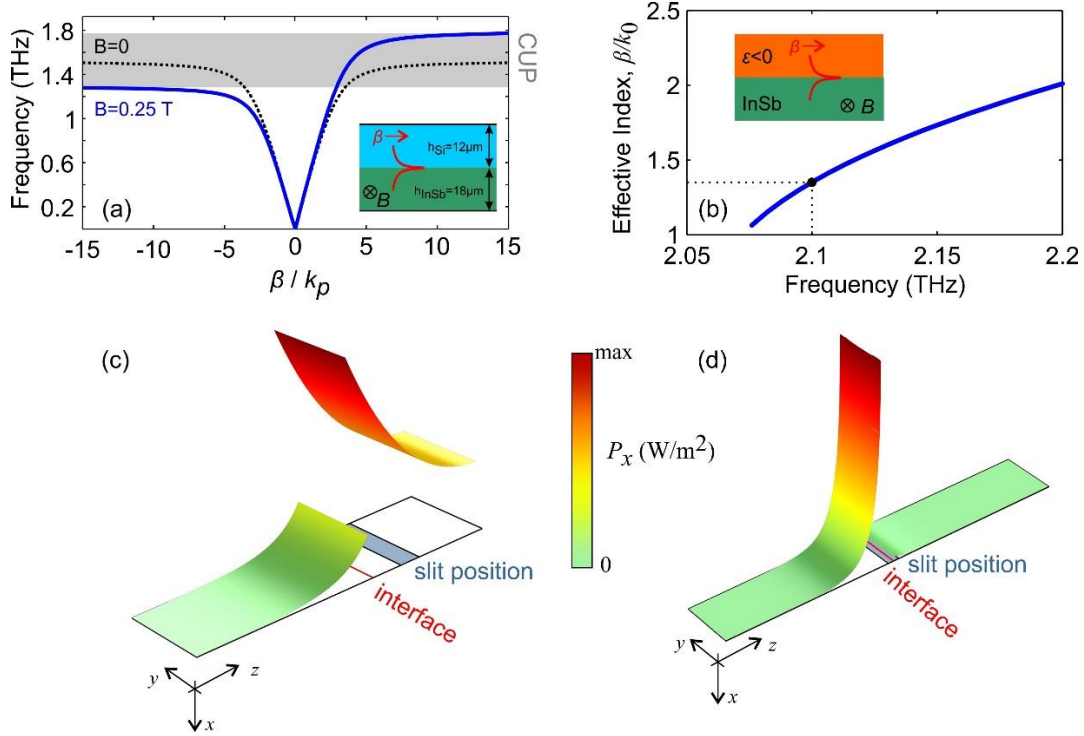

**Fig. 2.** (**a**) Band-diagram of the first structure, where $\beta$ is the guide's longitudinal propagation constant. The permittivity of the magnetized semiconductor (InSb) layer is a $3 \times 3$ tensor, owing to the applied magnetic biasing ($B$), and is given by $\varepsilon_{\mathrm{InSb}} = \varepsilon_0 \varepsilon_{\mathrm{inf}} \left[ \varepsilon_1(B) \; 0 \; i\varepsilon_2(B); \; 0 \; \varepsilon_3 \; 0; \; -i\varepsilon_2(B) \; 0 \; \varepsilon_1(B) \right]$, where $\varepsilon_1 = 1 - (\omega + i\nu)\omega_p^2/\{\omega[(\omega + i\nu)^2 - \omega_c^2]\}$, $\varepsilon_2 = \omega_c \omega_p^2/\{\omega[(\omega + i\nu)^2 - \omega_c^2]\}$, and $\varepsilon_3 = 1 - \omega_p^2/[\omega(\omega + i\nu)]\}$. For InSb it is $\varepsilon_{\mathrm{inf}} = 15.6$, $\omega_p = 4\pi \times 10^{12}$ rad/s, $\omega_c = \omega_c(B) = 0.25\omega_p$ is the cyclotron frequency, while the loss factor is $\nu = 5 \times 10^{-3}\omega_p$. The CUP frequency region is highlighted in gray. (**b**) Effective refractive index $n_{\mathrm{eff}} = \beta/k_0$ of the second structure vs. frequency. The parameters of the plasmonic superstrate are $\omega_p' = 2\omega_p$ (thus possessing a negative permittivity as long as $\omega_0 < 2\omega_p$), $\varepsilon_{\mathrm{inf}} = 15.6$, $\nu = 0$. Here, the unidirectional region extends between approximately $1.03\omega_p - 1.1\omega_p = 2.06 - 2.2$ THz, and is the only frequency region shown in the figure. For the results of Fig. 3(b) we operated at $\omega_0 = 1.05\omega_p = 2.1$ THz ($n_{\mathrm{eff}} = 1.3525 - \mathrm{j}0.0478$). Note that the magnetic field is $-0.25$T in this case (negative sign) because now the layer above InSb possesses a negative permittivity. (**c**), (**d**) Guided power profile distributions ($P_x$ component) for the structure shown, respectively, in the insets of (**a**), (**b**). From (**c**) we note that most of the power resides in the Si superstrate region, whereas for the structure of (**d**) most of the power resides in the InSb substrate region. The position of the materials interface and a possible slit position are clearly marked.

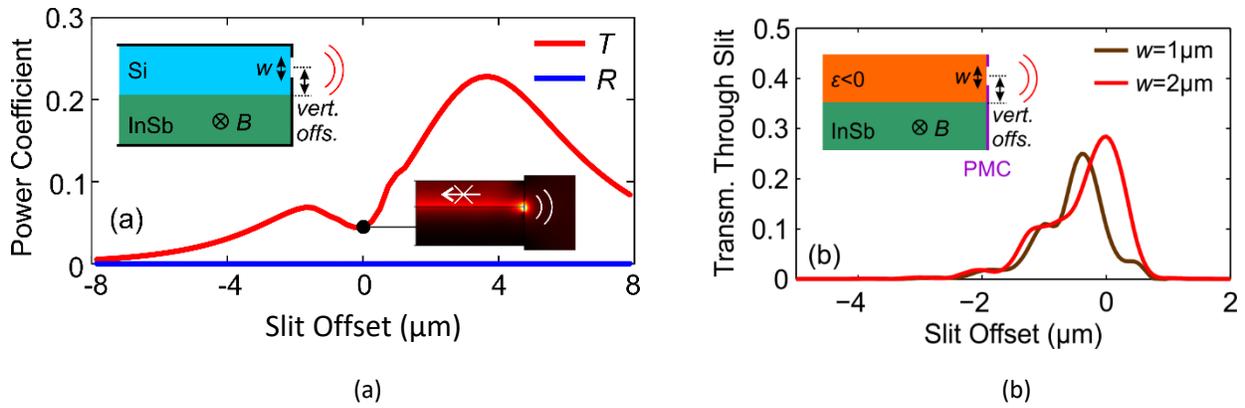

(a)

(b)

**Fig. 3.** (**a**) Transmission efficiency (power coefficient) through the slit of the first structure, for various slit offsets and for incident frequency $f$ = 1.4 THz. For a slit width of 2 μm (= $\lambda_0/107$), the maximum transmission through the slit is observed for a vertical offset of ~3.75 μm and equals ~23%. The slit width and offset definitions are marked in the upper inset. Note that in all cases the reflection coefficient is zero throughout, owing to the unidirectional nature of the mode; this is also seen in the field plot (lower inset) by the absence of a standing wave in the waveguide segment. (**b**) Transmission efficiency through the slit of the second structure as a function of vertical offset for an incident $f$ = 2.1 THz wave. For a slit width of 1 μm ($\lambda_0/138$; $\lambda_{eff}/72$), transmission through the slit reaches 25% at a slit offset of -0.4 μm (black line). If the slit width becomes 2 μm ($\lambda_0/69$; $\lambda_{eff}/36$), transmission reaches 28.5% for an offset of 0 μm (red curve).

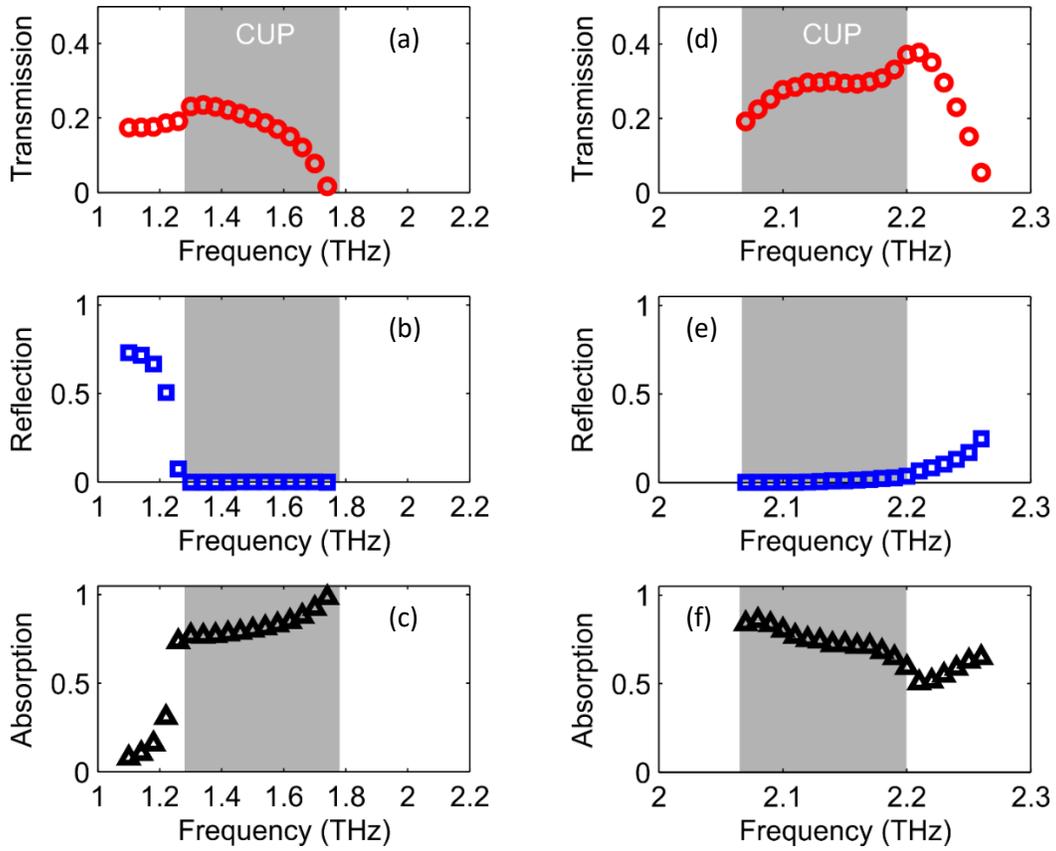

**Fig. 4.** Transmission, reflection and absorption coefficients, inside and near the 'complete unidirectional propagation' (CUP) band (grey area), for a 2 μm slit of: (**a-c**) the unidirectionally-guiding (non-topological) structure; and (**d-f**) the unidirectionally-guiding topological structure – for the optimum respective slit positions in both cases. Note that in both cases the reflection coefficient inside the CUP band is virtually zero as it should be from theory.